\documentclass[12pt,preprint]{aastex}
\shorttitle{Cooling of Accelerated Nucleons in GRBs}
\shortauthors{Asano}
\begin{document}
\title{Cooling of Accelerated Nucleons and Neutrino Emission in Gamma-Ray Bursts}
\author{Katsuaki Asano}
\affil{Division of Theoretical Astronomy, National Astronomical Observatory of Japan,
              2-21-1 Osawa Mitaka Tokyo, Japan}

\email{asano@th.nao.ac.jp}

\begin{abstract}
Using Monte Carlo simulations, we demonstrate
photopion production from Fermi-accelerated protons
and the resulting neutrino production in gamma-ray bursts.
Unless internal shocks occur
at quite large distance from the center,
ultra high-energy protons are depleted by photopion
production and synchrotron radiation.
Internal shocks at fiducial distance cause
neutrino bursts, which accompany gamma-ray
bursts originating from electromagnetic cascades.
\end{abstract}


\section{INTRODUCTION}

\indent

The rapid time variabilities and the compactness problem
(see, e.g., a review by Piran 1999) suggest that
gamma-ray bursts (GRBs) should arise from
internal shocks within relativistic flows.
In the standard model a strong magnetic field
is generated, and electrons are Fermi-accelerated
in shocked regions.
Because of the high Lorentz factor of the flows,
synchrotron radiation by accelerated electrons
is blue-shifted, and observed as gamma-ray photons.

The physical conditions in the shocked region imply
\citep{wax95} that protons may be also Fermi-accelerated
to energies $>10^{20}$ eV.
If the total energy of protons, accelerated over the energy range $10^{19}$-$10^{21}$ eV,
is comparable to the energy of the gamma-rays,
it can explain the flux of ultra-high-energy cosmic rays (UHECRs).
This is an attractive idea for the cosmic-ray source.

High energy protons in the GRB photon field can create high-energy
neutrinos via photopion production \citep{wax97,wax99,gue01,der03,gue04}.
If the above prediction is true,
a $1 {\rm km}^2$ neutrino detector may detect neutrinos
correlated with GRBs in the near future \citep{wax97,wic04}.
However, if the proton photopion ``optical depth'' is much higher
than unity, protons lose their energies in GRB sources.
In this case GRBs cannot be the sources of UHECRs.
\citet{asa03} (hereafter AT03) analytically show that
high-energy protons cool rapidly via photopion production
for fiducial parameters in GRB physics
and that most of the proton energy converts to photons via
$\pi^0$ decay, pion synchrotron, muon synchrotron, and
the electromagnetic cascade of electron-positron pairs.

In this paper, using the Monte Carlo method,
we simulate numerically the proton cooling and resulting
neutrino emission in order to confirm quantitatively whether
GRBs can be sources of UHECRs or not.
Our method pursues energy loss processes of each nucleon
interacting with photons and the magnetic field
during the dynamical timescale.
Therefore, we can treat energy loss processes correctly
even in the case that a nucleon creates pions multiple times
in the dynamical timescale.
\citet{der03} carried out similar calculations
using a different method from ours.
Our purpose is to obtain a physical condition to produce UHECRs,
while \citet{der03} mainly discussed neutrino production.
In \S 2 we explain our model and method.
The physical situation we consider is clearly shown.
The numerical results are in \S 3.
An analytically detailed description of the results can be found in AT03.
Our conclusions are summarized in \S 4.

\section{MODEL AND METHOD}

We simulate nucleon cooling in the GRB photon field
by the Monte Carlo method.
We adopt a relatively low GRB energy and a small magnetic field
for UHECR production to be as advantageous as possible.
The situation we consider is as follows.

In the standard model gamma-ray photons are emitted
from shocked shells moving with the bulk Lorentz factor $\Gamma >100$
(see, e.g., a review by Piran 1999).
In most cases
the observed isotropic energies of GRBs are larger than $10^{52}$ ergs \citep{fra01},
and the number of pulses per burst is less than $10$ \citep{mit98}.
According to the above facts,
we consider one shell of $\Gamma=300$, in which gamma-ray photons are produced,
and the total energy of those photons is $E_\gamma=10^{51}$ ergs
in spherical symmetry.
The shell width is constrained by the observed
variability time-scale, $\delta t > 1$ ms, and the duration time,
$\sim 10$ s.
Therefore, we assume that the shell width in the shell rest frame
$l$ is in the range of $10^{10}$-$10^{14}$ cm.
This implies the average luminosity $L=9 \times 10^{49}$-$9 \times 10^{53}$
ergs ${\mbox s}^{-1}$.
The photon energy density $U_\gamma=E_\gamma/(4 \pi \Gamma R^2 l)$
in the shell rest frame depends on the radius $R \sim 
\Gamma^2 c \delta t$ where an internal shock occurs.
In order to accelerate protons to energies $>10^{20}$ eV,
the internal shocks should occur at radii $\geq 10^{13}$ cm \citep{wax95}.
Therefore, we consider cases with $R \geq 10^{13}$ cm only.
The energy density of magnetic field $U_B=B^2/8 \pi$ is assumed to be $0.1 U_\gamma$.

In our simulation, adopting the typical observed GRB spectra,
the photon number density 
in the energy range $\epsilon_\gamma+d \epsilon_\gamma$
in the shell rest frame is set at $n(\epsilon_\gamma)\propto \epsilon_\gamma^{-1}$
for 1 eV $<\epsilon_\gamma< $ 1 keV
and $\epsilon_\gamma^{-2.2}$ for
1 keV $<\epsilon_\gamma< 10$ MeV.
The break energy 1 keV corresponds to 300 keV in the observer frame.
For $\epsilon_\gamma<1$ eV the synchrotron self absorption may be crucial
\citep{gra00}, while the pair absorption may be crucial for $\epsilon_\gamma>10$ MeV
(e.g., see AT03).
Thus, we neglect photons in such energy regions.

Since GRBs are possible sources of UHECRs
with enegies $>10^{20}$ eV,
we assume Fermi accelerated protons are injected in the
energy range of $\epsilon_{\rm p}=10^{10}$-$10^{19}$ eV in the shell
rest frame as $\dot{N}(\epsilon_{\rm p}) \propto \epsilon_{\rm p}^{-2}$.
Although the lowest energy of Fermi-accelerated particles
is unknown,
it is natural to consider that accelerated protons are distributed
to quite a lower energy.

Relativistic protons ($\epsilon_{\rm p}=\gamma_{\rm p} m_{\rm p} c^2
\gg m_{\rm p} c^2$) in the GRB photon field create
pions and lose their energies via photopion production.
In the shell rest frame the photon distribution is isotropic.
Then, the time scale of photopion production, $t_{\pi}$,
is written as
\begin{eqnarray}
t_{\pi}^{-1}(\gamma_{\rm p})=2 \pi c \int_{-1}^1 d\mu
(1-\beta_{\rm p} \mu)
\int d\epsilon_\gamma
\frac{n(\epsilon_\gamma)}{4 \pi}
\sigma_\pi(\chi),
\label{tau}
\end{eqnarray}
where $\beta_{\rm p}=\sqrt{\mathstrut 1-1/\gamma_{\rm p}^2} \simeq 1$
and $\mu$ is the cosine of the photon incident angle.
From \citet{ste68}, based on experiments,
we approximate the photopion production cross section
by a broken power-law profile as
$\sigma_{\pi}(\chi) =5 \times 10^{-28} (\chi/590)^{3.2} {\rm cm^2}$ 
for $290<\chi<590$ and 
$\sigma_{\pi}(\chi) =5 \times 10^{-28} (\chi/590)^{-0.7} {\rm cm^2}$ 
for $590<\chi<9800$, where $\chi m_{\rm e} c^2$
is the photon energy in the proton rest frame.
The energy of a photon is expressed as 
$\epsilon_\gamma=\chi m_{\rm e} c^2/\gamma_{\rm p} (1-\beta_{\rm p} \mu)$.

Secondary neutrons produced in the channel with isospin flip,
$p \gamma \to n \pi^+$, also create pions on almost
the same time-scale as $t_{\pi}$.
We use equation (\ref{tau}) for both protons and neutrons.
In the $\Delta$-approximation \citep{ste73}
the creation ratio $\pi^\pm : \pi^0$ is $1 :2$.
In the power-law target photon field, however,
the ratio $\pi^\pm : \pi^0$ is close to $2 :1$ \citep{rac98},
because the charged pion production rate increases
away from the $\Delta$-resonance.
So, we adopt $\pi^\pm : \pi^0 = 2 :1$ in this simulation.
For simplicity the nucleon inelasticity is fixed as
$K \equiv \Delta \epsilon_{\rm p}/\epsilon_{\rm p}=
\Delta \epsilon_{\rm n}/\epsilon_{\rm n}=0.2$,
and the process of multiple pion production,
which enhances the nucleon cooling efficiency, is neglected.

Extremely high energy protons can cool via synchrotron radiation with
the power per unit photon energy
\begin{eqnarray}
\frac{dW}{dt d\epsilon_\gamma}=\frac{\sqrt{\mathstrut 3} e^3 B \sin{\alpha}}
{h m_{\rm p} c^2} F(x),
\end{eqnarray}
where $x=2 m_{\rm p} c \epsilon_\gamma/(3 \gamma_{\rm p}^2 \hbar e B \sin{\alpha})$
and $F(x)$ is the synchrotron function \citep{ryb79}.
The pitch angle $\alpha$ is determined from random numbers
in our Monte Carlo simulation.
The cooling time scale
due to synchrotron radiation is
$t_{\rm SY}=3 (\gamma_{\rm p}-1) m_{\rm p}^3 c^5/2 e^4 B^2 \gamma_{\rm p}^2
\beta_{\rm p}^2
\sin^2{\alpha}$.

Another possible mechanism of the proton cooling is inverse Compton (IC) scattering.
The scattering rate is obtained from the Klein-Nishina cross section
$\sigma_{\rm KN}$ \citep{ryb79} by exchanging mass $m_{\rm e} \to m_{\rm p}$.
In the proton rest frame, when an photon with energy $\epsilon_0'$
is scattered, the energy of the scattered photon is
\begin{eqnarray}
\epsilon_1'=\frac{\epsilon_0'}{1+\frac{\epsilon_0'}{m_{\rm p} c^2}
(1-\mu_{\rm s})},
\end{eqnarray}
where $\mu_{\rm s}$ is the cosine of the scattering angle.
The probability distribution of the scattering angle is
\begin{eqnarray}
P_{\mu} \propto \frac{\epsilon_1'}{\epsilon_0'^2} \left(
\frac{\epsilon_0'}{\epsilon_1'}+\frac{\epsilon_1'}{\epsilon_0'}
-(1-\mu_{\rm s}^2) \right).
\label{P}
\end{eqnarray}
The Lorentz transformation yields the scattered photon energy
$\epsilon_1$ in the shell rest frame.
The cooling time-scale due to IC is
\begin{eqnarray}
t_{\rm IC}^{-1}(\gamma_{\rm p})=
\frac{1}{(\gamma_{\rm p}-1) m_{\rm p} c^2}
2 \pi c \int_{-1}^1 d\mu (1-\beta_{\rm p} \mu)
\int d\epsilon_\gamma (\epsilon_1-\epsilon_\gamma)
\frac{n(\epsilon_\gamma)}{4 \pi}
 \sigma_{\rm KN}(\epsilon_0'),
\end{eqnarray}
where $\epsilon_0'=\gamma_{\rm p} \epsilon_\gamma
(1-\beta_{\rm p} \mu)$.
To obtain $t_{\rm IC}$, the numerical integration
is carried out with $\mu_{\rm s}$ determined
from equation (\ref{P}).
This numerical integration process also gives the IC spectra
at once.

We pursue $2.7 \times 10^4$ nucleons in each computation.
Given a nucleon energy,
the cooling time scales $t_\pi$, $t_{\rm SY}$, and $t_{\rm IC}$
are determined.
Let us denote the cooling time scale due to photon emission by $t_\gamma$,
which satisfies $t_\gamma^{-1}=t_{\rm SY}^{-1}+t_{\rm IC}^{-1}$.
Assuming that nucleons interact with photons and the magnetic field
during the dynamical time scale $t_{\rm d}=l/c$,
we pursue the nucleon cooling process by a time step
$\Delta t=\min(t_\pi, t_\gamma)/10$ and $t_\pi/10$
for protons and neutrons, respectively.
The time step will change as nucleons lose their energies.
The pion production probability in each time step
is $1-\exp(-\Delta t/t_\pi)$.
If nucleons create charged pions, the species of the nucleons
is changed ($p \leftrightarrow n$).
After $t_{\rm d}$ s have passed, nucleons are considered to escape from the shell.
The neutron-decay emissions will produce nonthermal emission
signatures.
However, we can distinguish those emissions from
the prompt burst,
since the flux of neutrinos and photons originated from
the neutron-decay is faint because of the long decay time-scale.
We do not discuss those emissions below.

\section{RESULTS}

\subsection{Proton cooling}

The time-scales $t_\pi$, $t_{\rm SY}$, and $t_{\rm IC}$
are proportional to $R^2 l$,
while the dynamical time-scale $t_{\rm d} \propto l$ (see AT03).
The proton cooling process is determined by the ratios
$\tau_{\pi}=t_{\rm d}/t_{\pi}$,
$\tau_{\rm SY}=t_{\rm d}/t_{\rm SY}$, and
$\tau_{\rm IC}=t_{\rm d}/t_{\rm IC}$,
which are proportional to $R^{-2}$.
Thus, the proton cooling process is independent of $l$.
Under the situation we consider, the proton cooling efficiency
depends on the total photon energy $E_\gamma$ and the radius $R$,
rather than on the luminosity $L \propto E_\gamma/l$.

In Figure 1 we plot the ratios $\tau_{\pi}, \tau_{\rm SY}, \tau_{\rm IC}$
for $R=10^{13}$ cm and $E_\gamma=10^{51}$ erg.
From the condition to create pions $\epsilon_{\rm p} \epsilon_\gamma
\geq 0.2 {\rm GeV}^2$,
the behavior of $\tau_\pi$ is explained as follows.
For $\Gamma \epsilon_{\rm p} <6 \times 10^{16}$ eV,
protons produce pions with photons whose energies are above the
break energy $\Gamma \epsilon_\gamma=300$ keV
[$n(\epsilon_\gamma)\propto \epsilon_\gamma^{-2.2}$].
In this case
the number of the interacting photons is proportional to
$\sim \epsilon_{\rm p}^{1.2}$,
which implies $\tau_{\pi} \propto \epsilon_{\rm p}^{1.2}$.
Above $6 \times 10^{16}$ eV protons interact with
photons in a low-energy region [$n(\epsilon_\gamma)\propto \epsilon_\gamma^{-1.0}$].
As a result, $\tau_{\pi}$ is nearly constant in this energy region.
It is easily confirmed that
the simple approximation in \citet{wax97},
which was adopted in simulations in \citet{gue01,gue04},
produces a close value to our numerical estimate, $\tau_{\pi}$.
Since \citet{wax97} adopted a very low energy,
$E_\gamma=L \delta t=10^{48}$ ergs for a shell,
the pion production efficiency in their case becomes lower than ours.
Beyond $6 \times 10^{19}$ eV, the threshold energy of target photons
to create pions is below the cut-off energy (0.3 keV in the observer frame).
As a result, $\tau_\pi$ decreases above this energy.

For $\Gamma \epsilon_{\rm p}>3 \times 10^{20}$ eV,
the proton synchrotron cooling dominates photopion production.
Although the photon energy density is larger than
the magnetic energy density ($U_B=0.1 U_\gamma$),
IC emission is negligible because of
the Klein-Nishina suppression.

Figure 1 indicates that all protons above $10^{15}$ eV may
lose their energies for $R=10^{13}$ cm.
The result of the numerical simulation shown in Figure 2
is consistent with the above discussion.
Nucleons with energies $>10^{15}$ eV are extinguished.
The energy protons lose is mainly transferred to pions and synchrotron
photons (see Table 1).
The highest energy protons ($>6 \times 10^{19}$ eV) cool via synchrotron radiation,
so that the pion distribution has a break at $\sim 10^{19}$ eV.
The low-energy tail of the synchrotron spectrum exhibits $n(\epsilon_\gamma)
\propto \epsilon_\gamma^{-1.5}$, as is usually seen in the fast cooling case \citep{gra00}.
The weak peak at $\sim 10^5$ eV in the synchrotron spectrum may be due to
low-energy protons whose photopion ``optical depth'' is small enough.
The IC spectrum is complex because of the Klein-Nishina suppression.

Even for $R=10^{13.5}$ cm (Fig. 3),
nucleons $>10^{16}$ eV are depleted by 2 orders of magnitude.
Figure 4 shows the case for $R=10^{14}$ cm.
In this case GRBs can be sources of UHECRs, although
nucleons $> 10^{20.5}$ eV are reduced by the synchrotron cooling.
The maximum value of $\tau_\pi$ is $\sim 1$ because of
the low photon density.
The highest energy protons create a few pions and
lose a few tens of percent of their energies.


\subsection{Neutrinos}

Using the same method, we simulate behavior of
charged pions and muons until they decay into positrons (electrons)
and neutrinos.
The life times of pions and muons, $T_\pi=\gamma_\pi T_{\pi,0}$
and $T_\mu=\gamma_\mu T_{\mu,0}$,
($T_{\pi,0}=2.6 \times 10^{-8}$ s
and $T_{\mu,0}=2.2 \times 10^{-6}$ s are the life times of pions
and muons at rest, respectively)
are independent of the shell width $l$.
Therefore, the energies charged pions and muons lose before their decays
depend on $l$ differently from the proton cooling.

When a pion decays $\pi^+ (\pi^-) \to \mu^+ \nu_\mu (\mu^- \bar{\nu_\mu})$,
the energy fraction a muon obtains
is $\sim m_\mu/m_\pi \sim 0.76$.
The rest of the energy goes to a neutrino.
We assume that the energy of a muon at its decay will be shared equally
by a positron (electron), a neutrino, and an anti-neutrino.
We neglect neutrinos due to neutron decay $n \to p e^- \bar{\nu_{\rm e}}$,
whose time-scale is much longer than $t_{\rm d}$.

Figure 5 shows the resulting spectra of neutrinos and photons
emitted by pions and muons for $R=10^{13}$ cm and $l=10^{10}$ cm.
Pions and muons above the energy where the life time
is comparable to the cooling time $t_\gamma$
or dynamical time $t_{\rm d}$ (Waxman and Bahcall 1997; AT03)
lose their energy before their decay.
For $R=10^{13}$ cm and $l=10^{10}$ cm,
one half of the initial energy of the pions is expended
via synchrotron radiation, and muons also lose their energy
via synchrotron ($\sim 35$\%) and IC ($\sim 17$\%) radiation.
Neutrinos are distributed as $n_\nu(\epsilon_\nu) \propto \epsilon_\nu^{-2}$
with a low- and high-energy cut-off.

The high-energy cut-off is determined by the life-time
and the cooling time of pions (muons).
Since a larger $l$ leads to a lower $U_B$,
the high-energy cut-off is roughly proportional to $l^{0.5}$
(see Fig. 6).
Wider shell decreases the contribution of IC photons
in comparison with synchrotron radiation,
because of the Klein-Nishina suppression.

The low-energy cut-off is determined
by the minimum energy of pions produced from protons,
which depends on only $R$, as was discussed in \S 3.1.
Both the low-energy and high-energy cut-off increase
roughly $\propto R$ (see Fig. 7).
If we observe neutrinos from GRBs,
the spectra give us parameters such as $l$ and $R$
independently of gamma-ray observation.

Figure 8 shows the ratio of the neutrino energy
to the initial proton energy ($\geq 10^{10}$ eV at injection).
For $R \leq 10^{14}$ cm, $\sim 10$\% of the proton energy
will be converted to a neutrino burst.
If we consider only muon neutrinos, which will be detected with a kilometer-scale
detector,
their energy fraction may be
a few percent of the accelerated protons.
This result is consistent with the prediction of \citet{der03}.

The neutrino energy is a part of the energy protons lose
via photopion production and synchrotron radiation.
The ``rest energy'' (the energies of neutral pions,
photons, electrons, and positrons)
will be converted to lower energy photons ($<10^9$ eV)
via electromagnetic cascades (see AT03),
because higher energy photons ($>10^9$ eV) will be absorbed
by electron-positron pair creation.
The total energy of such photons is quite larger than
the neutrino energy (see Fig. 9).
If the energy of Fermi-accelerated protons above $3 \times 10^{19}$ eV
is comparable to the GRB photon energy $E_\gamma$,
the energy of protons $>10^{10}$ eV is $4.5 E_\gamma$.
Thus, the energy of protons may be 1-5 $E_\gamma$.
In any case the energy of nucleons above $3 \times 10^{19}$ eV
will be lost for $R<10^{14}$ cm.
As long as the energy of UHECRs is assumed to be
comparable to the GRB energy,
gamma-ray photons originating from
protons may be a large contiribution to the GRB photons.

\section{CONCLUSIONS AND DISCUSSION}

Our simulations have shown that internal shocks should occur
at radii $\geq 10^{14}$ cm to generate UHECRs,
although siginificant GRB pulses have shorter time-scales of $<1$ s
\citep{nor96}.
On the other hand,
neutrino bursts are expected
for $R<10^{14}$ cm, while ultra-high-energy protons
are depleted.
The energy fraction of the neutrinos and neutrino spectra
have been calculated for various $R$ and $l$.
Neutrino spectra give us information on the physical
condition of GRBs,
although only the most powerful bursts, which are brighter than $10^{53}$ erg
or occur at $z< \sim 0.1$, produce detectable neutrino bursts
with a kilometer-scale detector \citep{der03}.

Smaller $E_\gamma$ and larger $R$ are favorable
to generate UHECRs.
At $R=10^{14}$ cm, the most probable shell width is
$l \sim R/\Gamma \sim 10^{12}$ cm.
In this case the magnetic field becomes $8 \times 10^3$ G
assuming $E_\gamma=10^{51}$ ergs and $U_B=0.1 U_\gamma$.
This strength seems rather weak to produce observed gamma-ray photons.
Since there are some unsolved problems and ambigious points
in the theory of GRB spectra \citep{pre98,tot99,asa03b},
we have simply adopted a typical observed spectra
without considering any physical conditions.
It is challeging to find a condition to make both UHECRs and
typical gamma-ray spectra be consistent.

To produce both UHECRs and a neutrino burst at the same time,
the radius is limitted to around $R \sim 10^{14}$ cm in our case.
Throughout this paper, we have adopted a
relatively small GRB energy $E_\gamma=10^{51}$ ergs as an optimistic case.
For brighter bursts ($E_\gamma>10^{51}$ ergs),
stronger magnetic fields ($U_B \sim U_\gamma$), and smaller
Lorentz factors ($\Gamma<300$),
internal shocks should occur at more distant radii to generate UHECRs.
The threshold radius behaves as roughly $\propto E_\gamma^{1/2}
\Gamma^{-1/2}$ under our assumption.
If protons are accelerated to very high energies
at both the internal shock phase and afterglow phase,
both UHECRs and neutrino burst may be produced.
In this case the neutrino bursts may be originated from
internal shocks, while most of the UHECRs are produced at the afterglow phase.

Throughout this paper the internal shock model has been applied.
However, the external shock model may be applicable for smooth
profile GRBs.
According to \citet{der02}, the neutrino flux from individual GRBs
is far too weak in the external shock model,
while neutrino bursts occur in the internal shock model
as shown in this paper.
In the external shock model,
ultra-high-energy neutrons are produced and travel
$\sim 100$ kpc before their numbers are depleted by $\beta$-decay.
Therefore, the external shock model predicts
detection of neutron $\beta$-decay halo emission at optical
and radio frequencies with luminosity $\sim 10^{35}$ ergs ${\rm s}^{-1}$.
On the other hand UHECRs may be extinguished for bright GRBs
in the internal shock model so that
neutron $\beta$-decay halos are not generated.
Future neutrino, optical, and radio observations
may provide tests of the two scenarios.
Measurement of ion composition of UHECRs from supernova remnants
with instruments such as {\it Auger} may be helpful
for investigating the origin of UHECRs.

Neutrino bursts surely accompany gamma-ray photons
originating from pion cascades.
Those photons may be not negligible
in comparison with the ``original GRB photons.''
In this case, as AT03 discussed,
GRB photons may arise from pion cascades,
rather than from initially existing electrons.

\acknowledgments

The author thanks the referee for useful comments.

\clearpage

\begin{figure}
\epsscale{.70}
\plotone{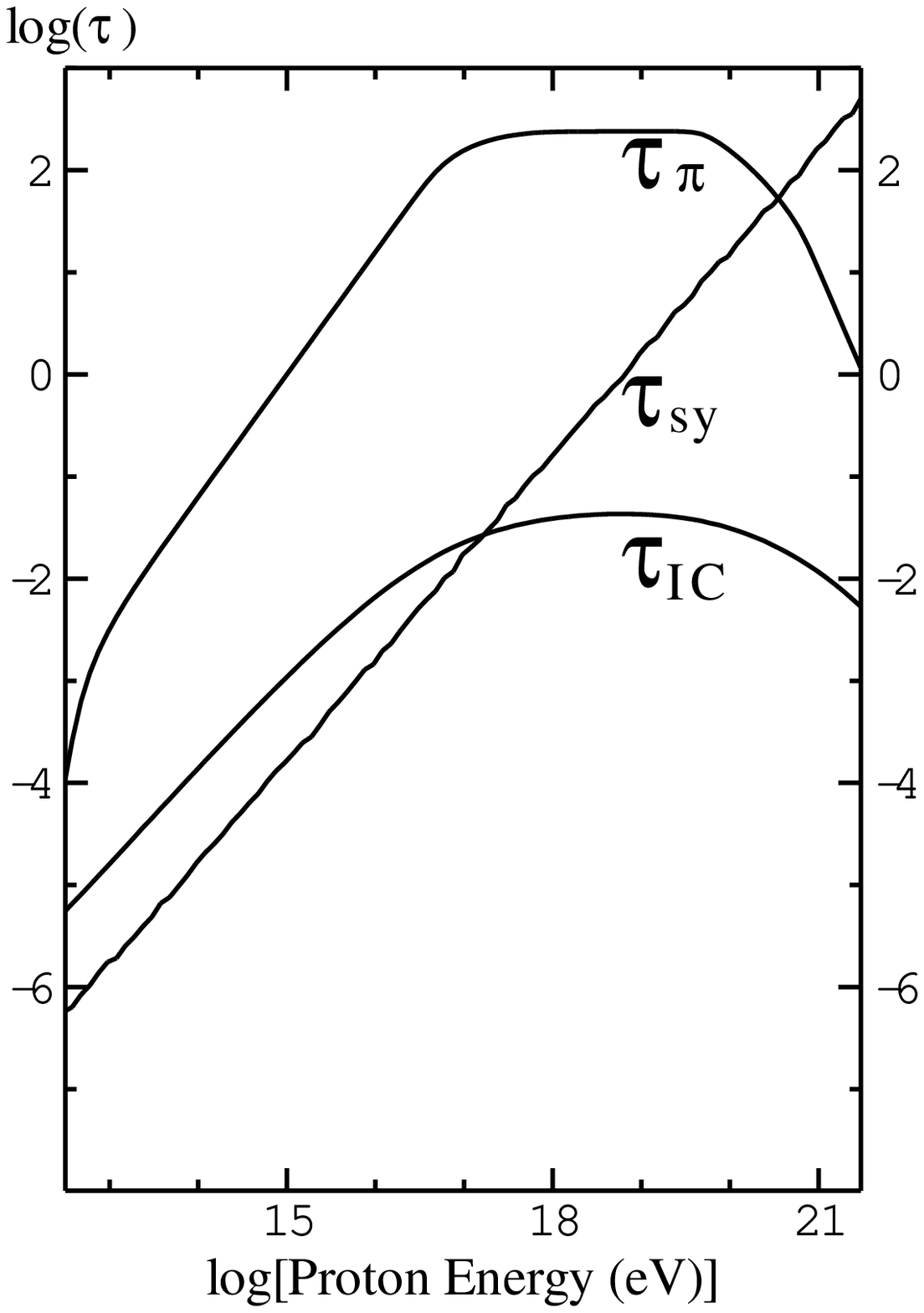}
\caption{Proton cooling efficiencies $\tau_{\pi}, \tau_{\rm SY}, \tau_{\rm IC}$
for $R=10^{13}$ cm and $E_\gamma=10^{51}$ ergs.
The proton energies are measured in the observer frame.}
\end{figure}

\clearpage

\begin{figure}
\plotone{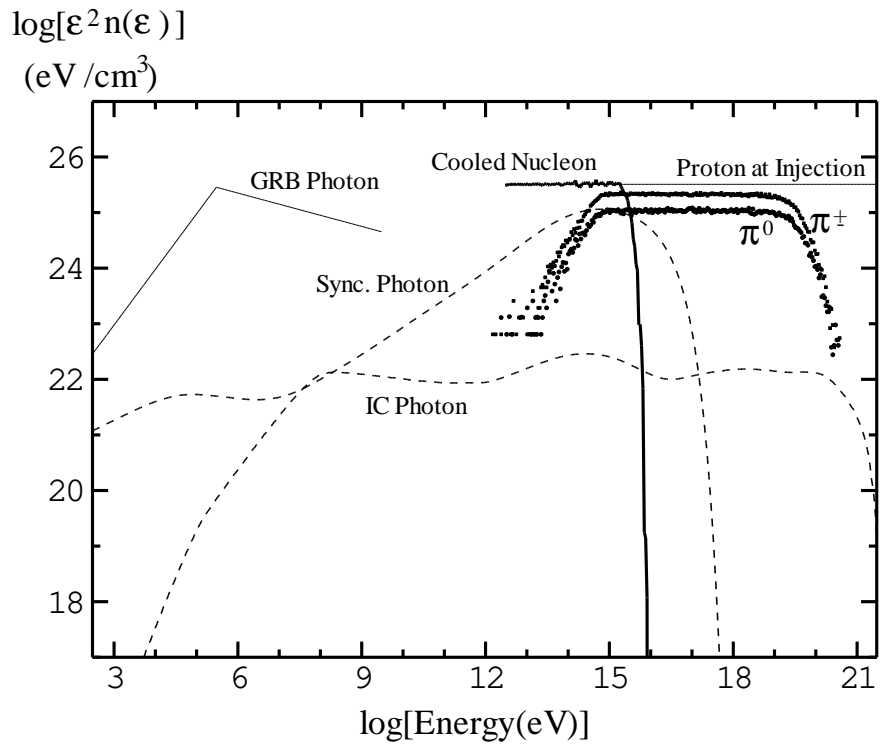}
\caption{Spectra of GRB photons, protons at injection,
created pions, photons emitted by protons ({\it dashed lines}),
and cooled nucleons (protons and neutrons) for $R=10^{13}$ cm.}
\end{figure}

\clearpage

\begin{figure}
\plotone{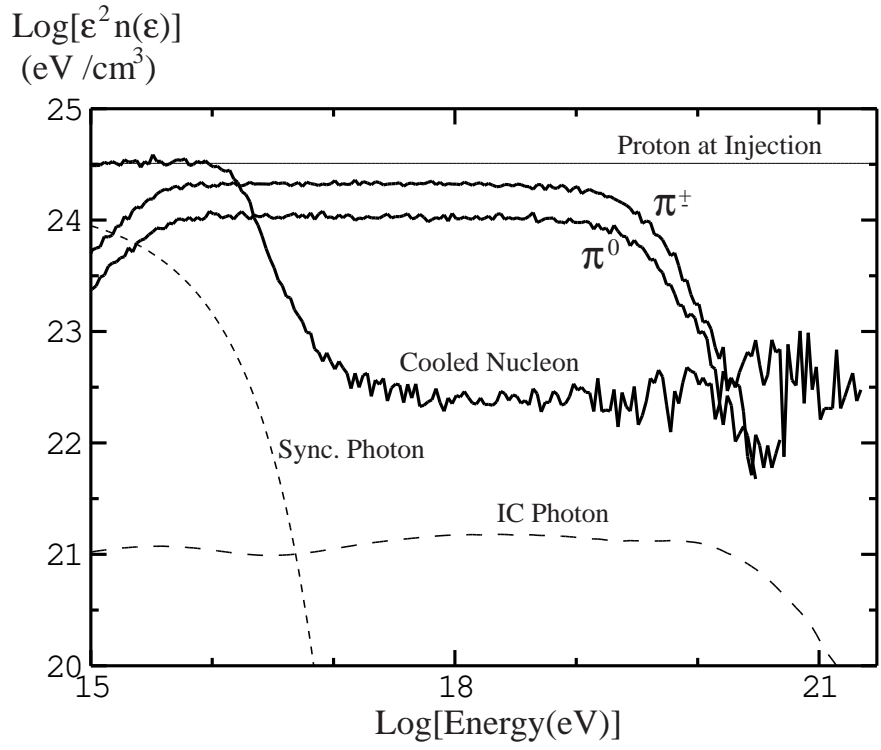}
\caption{Same as Fig. 2, but for $R=10^{13.5}$ cm.}
\end{figure}

\clearpage

\begin{figure}
\plotone{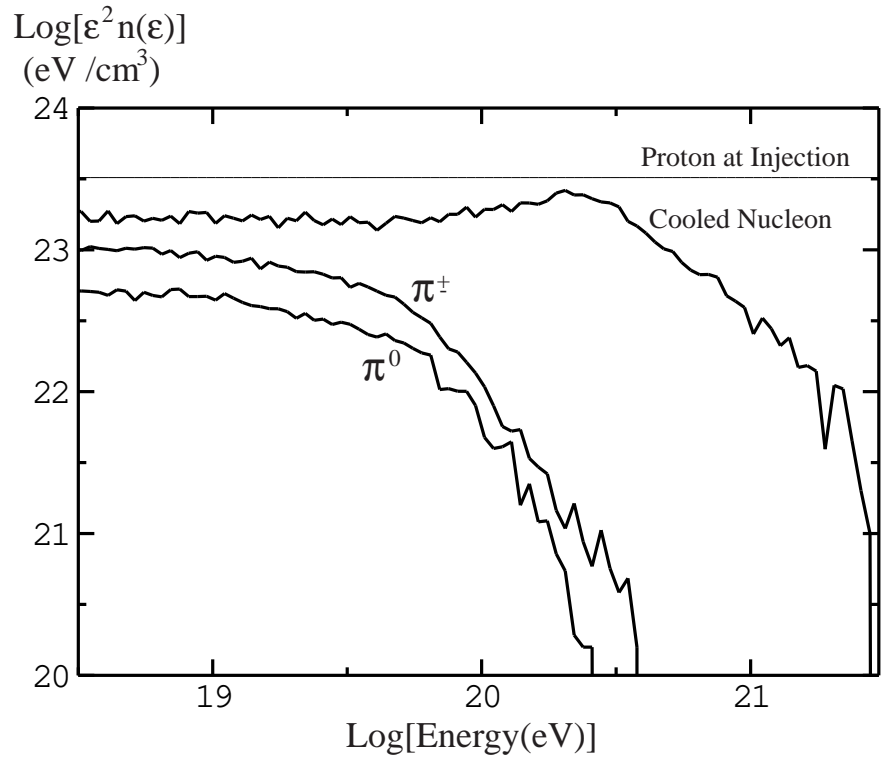}
\caption{Same as Fig. 2, but for $R=10^{14}$ cm.}
\end{figure}

\clearpage

\begin{figure}
\plotone{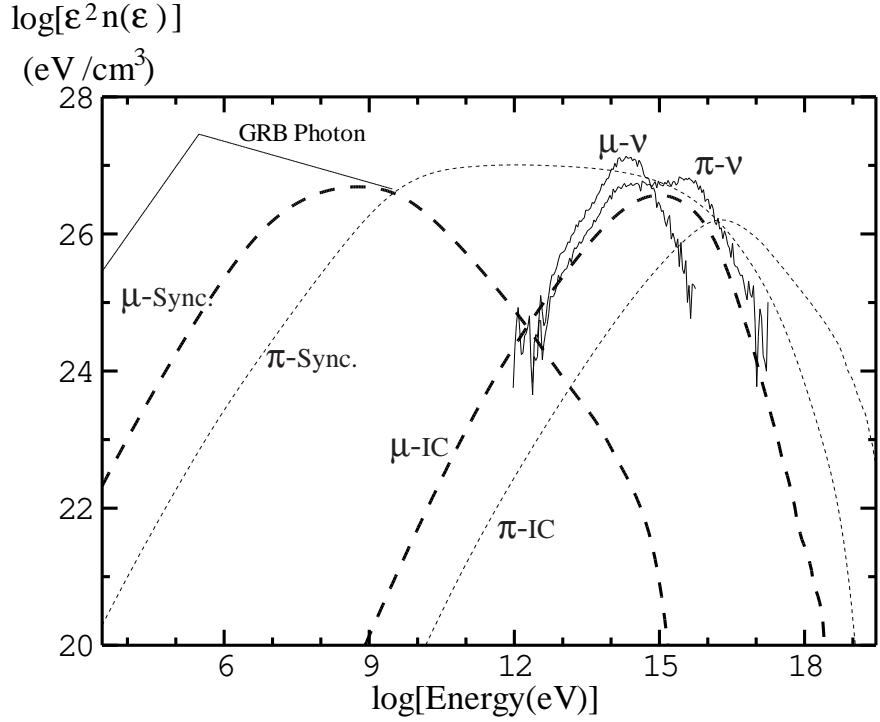}
\caption{Spectra of neutrinos ({\it solid lines})
from pion decay ($\nu_\mu$ and $\bar{\nu_\mu}$)
and muon decay ($\nu_{\rm e}$, $\bar{\nu_{\rm e}}$, $\nu_\mu$ and $\bar{\nu_\mu}$)
and synchrotron and IC photons emitted by pions ({\it thin dashed lines})
and muons ({\it thick dashed lines}) for $R=10^{13}$ cm and $l=10^{10}$ cm.}
\end{figure}

\clearpage

\begin{figure}
\plotone{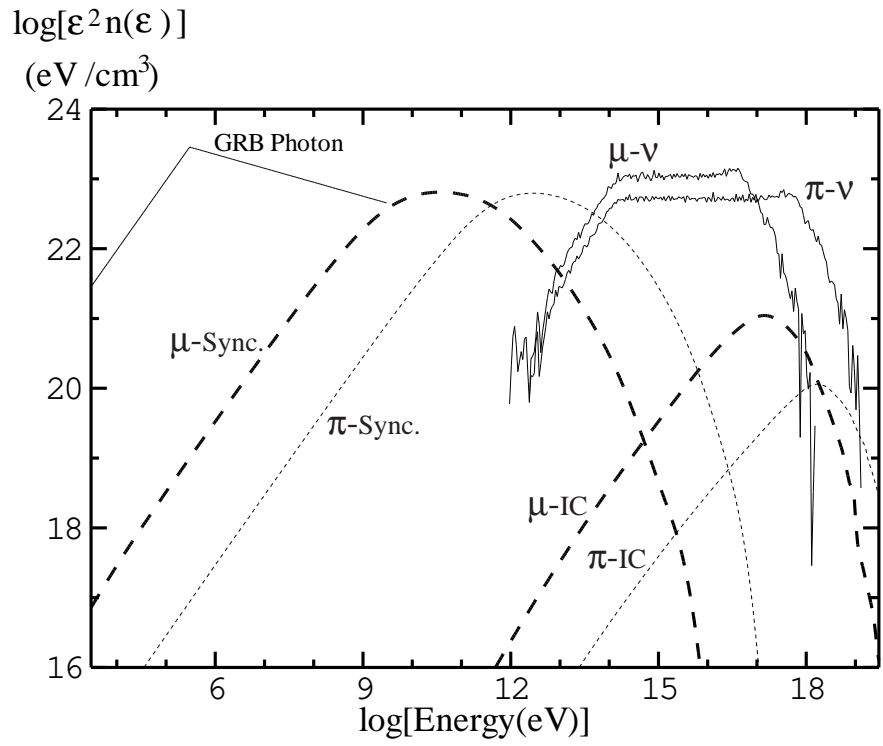}
\caption{Same as Fig. 5, but for $l=10^{14}$ cm.}
\end{figure}

\clearpage

\begin{figure}
\plotone{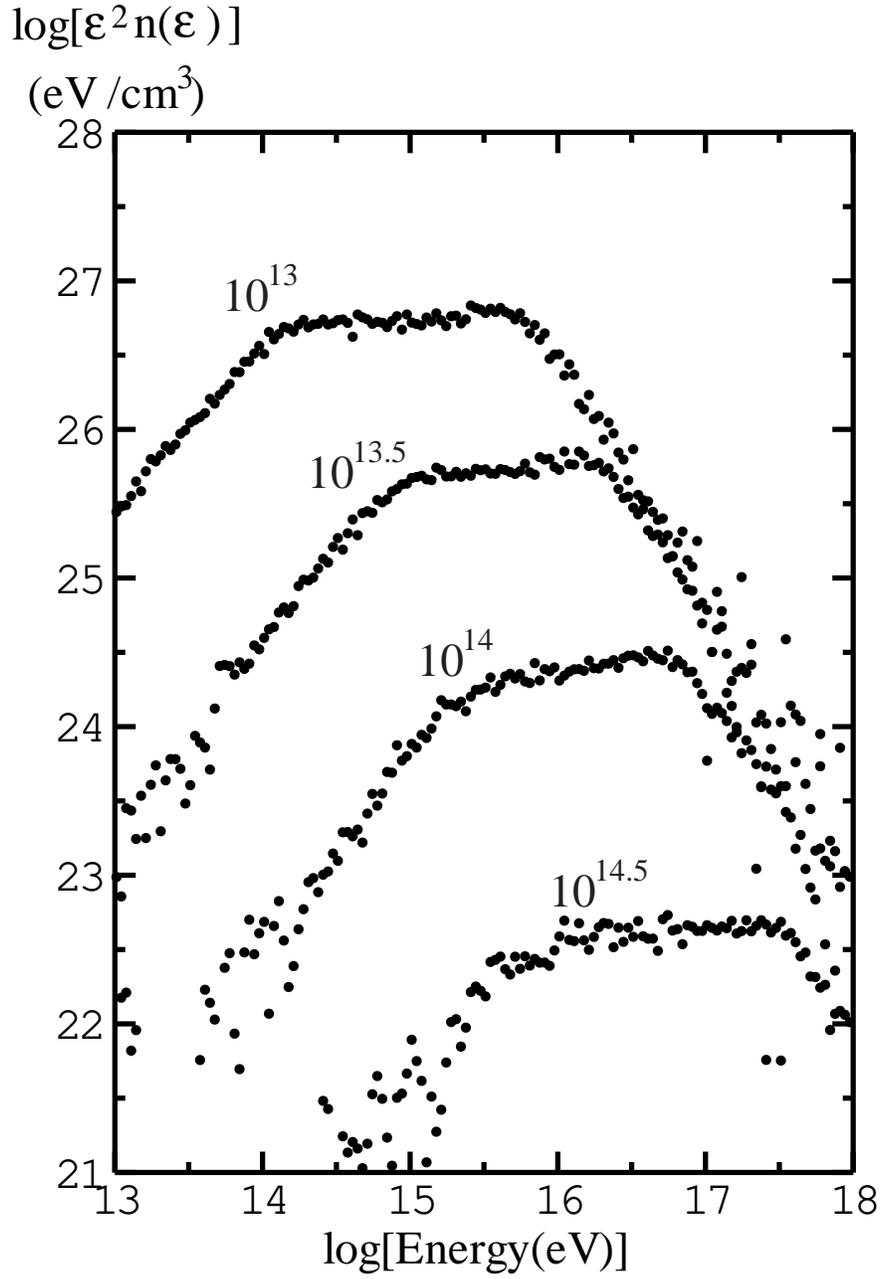}
\caption{Spectra of neutrinos from pion decay for $l=10^{10}$ cm.
The radius $R$ changes from $10^{13}$ to $10^{14.5}$ cm.}
\end{figure}

\clearpage

\begin{figure}
\plotone{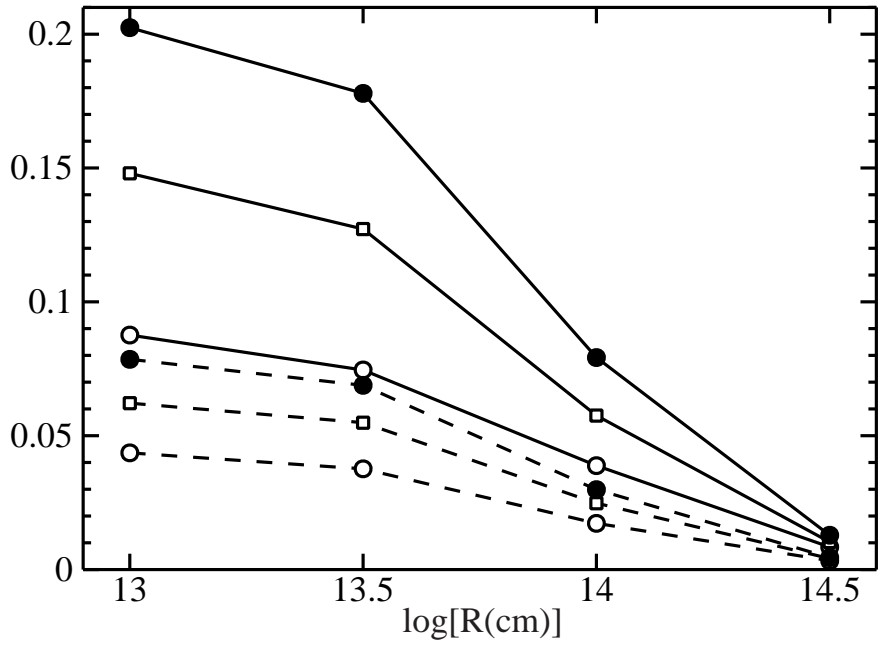}
\caption{Ration of neutrino energy to the initial proton energy.
Dashed lines are for neutrinos originated from pions only, and solid
lines are the total fraction.
The shell width is assumed to be $10^{10}$ ({\it open circles}),
$10^{12}$ ({\it rectangles}), and $10^{14}$ ({\it filled circles}) cm,
respectively.}
\end{figure}

\clearpage

\begin{figure}
\plotone{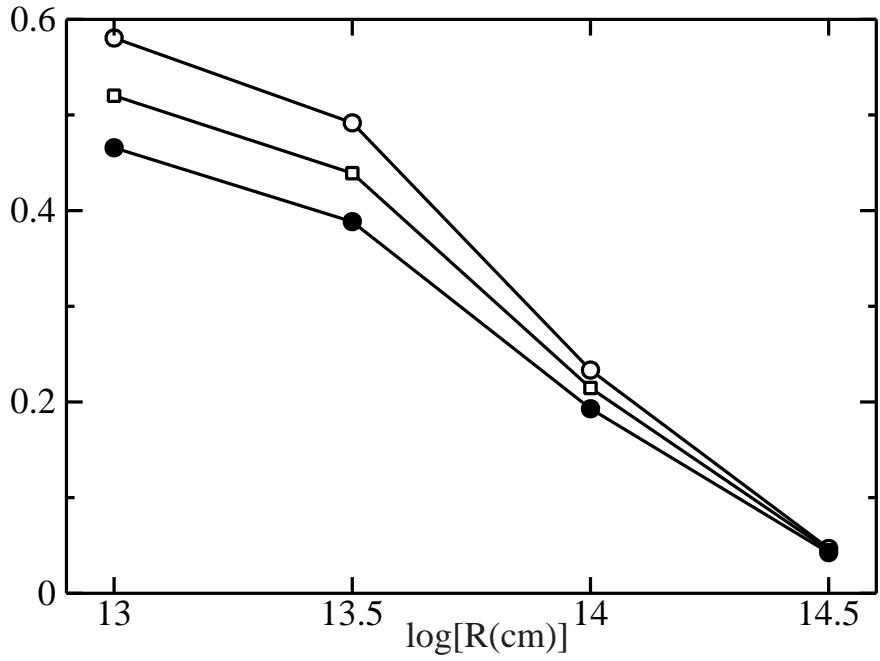}
\caption{Ratio of ``rest energy'' (see text),
which will be converted to gamma-ray photons,
to the initial proton energy.
The shell width is assumed to be $10^{10}$ ({\it open circles}),
$10^{12}$ ({\it rectangles}), and $10^{14}$ ({\it filled circles}) cm, respectively.}
\end{figure}

\clearpage

\begin{table}
\begin{center}
\caption{Energy Fraction Each Species of Particle Contributes
to the Total Energy of Protons at Injection.}
\begin{tabular}{lccc}
\hline \hline
$R$ (cm) & $N$ & $\pi$ & $\gamma$   \\ \hline
$10^{13}$ & 0.33 & 0.57 & 0.10 \\
$10^{13.5}$ & 0.43 & 0.47 & 0.10 \\
$10^{14}$ & 0.73 & 0.19 & 0.08 \\
$10^{14.5}$ & 0.94 & 0.03 & 0.03 \\
$10^{15}$ & 0.99 & 0.002 & 0.01 \\ \hline
\end{tabular}
\tablecomments{
The notations $N$, $\pi$, and $\gamma$ represent
cooled nucleons, created $\pi^0$ and $\pi^\pm$, and
photons emitted directly by protons, respectively.}
\end{center}
\end{table}


\clearpage
\begin{thebibliography}{}
\bibitem[Asano and Kobayashi(2003)]{asa03b}
Asano, K., \& Kobayashi, S. 2003, \pasj, 55, 579
\bibitem[Asano and Takahara(2003)]{asa03}
Asano, K., \& Takahara, F. 2003, \pasj, 55, 433 (AT03)
\bibitem[Dermer (2002)]{der02}
Dermer, C. D. 2002, \apj, 574, 65
\bibitem[Dermer and Atoyan (2003)]{der03}
Dermer, C. D., \& Atoyan, A. 2003, Phys. Rev. Lett., 91, 071102
\bibitem[Frail et al.(2001)]{fra01}
Frail, D. A., Kulkarni, S. R., Sari, R., Djorgovski, S. G., Bloom, J. S.,
Galama, T. J., Reichart, D. E., Berger, E., Harrison, F. A., Price, P. A.,
Yost, S. A., Diercks, A., Goodrich, R. W., \& Chaffee, F. 2001, \apjl, 562, L55
\bibitem[Granot et al.(2000)]{gra00}
Granot, J., Piran, T., \& Sari, R. 2000, \apjl, 534, L163
\bibitem[Guetta et al.(2004)]{gue04}
Guetta, D., Hooper, D., Alvarez-Mu\~niz, J., Halzen, F.,
\& Reuveni, E. 2004, Astroparticle Phys., 20, 429
\bibitem[Guetta et al.(2001)]{gue01}
Guetta, D., Spada, M., \& Waxman, E. 2001b, \apj, 559, 101
\bibitem[Mitrofanov et al.(1998)]{mit98}
Mitrofanov, I. G., Pozanenko, A. S.,
Briggs, M. S., Paciesas, W. S., Preece, R. D., Pendleton, G. N.,
\& Meegan, C. A. 1998, \apj, 504, 925
\bibitem[Norris et al.(1996)]{nor96}
Norris, J. P., Nemiroff, R. J.,
Bonnell, J. T., Scargle, J. D., Kouveliotou, C., Paciesas, W. S.,
Meegan, C. A., \& Fishman, G. J. 1996, \apj, 459, 393
\bibitem[Piran (1999)]{pir99}
Piran, T. 1999, Phys. Rep., 314, 575
\bibitem[Preece et al.(1998)]{pre98}
Preece, R. D., Briggs, M. S, Mallozzi, R. S., Pendleton, G. N.,
Paciesas, W. S., \& Band, D. L. 1998, \apjl, 506, L23
\bibitem[Rachen and M\'esz\'aros (1998)]{rac98}
Rachen, J. P., \& M\'esz\'aros, P. 1998, Phys. Rev. D, 58, 123005
\bibitem[Rybicki and Lightman (1979)]{ryb79}Rybicki, G. B., \& Lightman, A. P. 1979,
Radiative Processes in Astrophysics (New York: Wiley-Interscience)
\bibitem[Stecker (1968)]{ste68}
Stecker, F. W. 1968, Phys. Rev. Lett., 21, 1016
\bibitem[Stecker (1973)]{ste73}
Stecker, F. W. 1973, \apss, 20, 47
\bibitem[Totani (1999)]{tot99}
Totani, T. 1999, \mnras, 307, L41
\bibitem[Waxman (1995)]{wax95}
Waxman, E. 1995, Phys. Rev. Lett., 75, 386
\bibitem[Waxman and Bahcall (1997)]{wax97}
Waxman, E., \& Bahcall, J. 1997, Phys. Rev. Lett., 78, 2292
\bibitem[Waxman and Bahcall (1999)]{wax99}
Waxman, E., \& Bahcall, J. 1999, Phys. Rev. D, 59, 023002
\bibitem[Wick et al.(2004)]{wic04}
Wick, S. D., Dermer, C. D., \& Atoyan, A. 2004, Astroparticle Phys., 21, 125
\end{thebibliography}
\end{document}